\documentclass[12pt]{article}
\usepackage{amsmath}
\def\be{\begin{equation}}
\def\ee{\end{equation}}
\def\arr{\begin{array}{rll}}
\def\ea{\end{array}}
\def\bea{\begin{eqnarray}}
\def\eea{\end{eqnarray}}

\def\N2{$N{=}2$}

\def\>{\rangle}
\def\<{\langle}
\def\+{\dagger}
\def\={\ =\ }

\begin{document}
\renewcommand{\thefootnote}{\fnsymbol{footnote}}
\begin{titlepage}
\setcounter{page}{0}
\vskip 1cm
\begin{center}
{\LARGE\bf  On the near horizon rotating black hole }\\
\vskip 0.5cm
{\LARGE\bf    geometries with NUT charges }\\
\vskip 2cm
$
\textrm{\Large Anton Galajinsky \ } \textrm{\Large and \ } \textrm{\Large Kirill Orekhov\ }
$
\vskip 0.7cm
{\it
Laboratory of Mathematical Physics, Tomsk Polytechnic University, \\
634050 Tomsk, Lenin Ave. 30, Russian Federation} \\
{E-mails: galajin@tpu.ru, orekhovka@tpu.ru}

\end{center}
\vskip 1cm
\begin{abstract} \noindent
The near horizon geometries are usually constructed by implementing a specific limit to
a given extreme black hole configuration. Their salient feature
is that the isometry group includes the conformal subgroup $SO(2,1)$. In this work, we turn the logic around and use the
conformal invariants for constructing Ricci--flat metrics in $d=4$ and $d=5$ where
the vacuum Einstein equations reduce to a coupled set of ordinary differential equations. In four dimensions the analysis
can be carried out in full generality and the resulting metric describes the $d=4$ near horizon Kerr--NUT black hole. In five dimensions we choose
a specific ansatz whose structure is similar to the $d=5$ near horizon Myers--Perry black hole.
A Ricci--flat metric involving five arbitrary parameters is constructed. A particular member of this family,
which is characterized by three parameters,
seems to be a natural candidate to describe the $d=5$ near horizon Myers--Perry black hole with a NUT charge.
\end{abstract}

\vspace{0.5cm}

\end{titlepage}

\renewcommand{\thefootnote}{\arabic{footnote}}
\setcounter{footnote}0

\noindent
{\bf 1.  Introduction}\\

\noindent
Motivated by the Kerr/CFT--correspondence \cite{str}, the near horizon black hole geometries in various dimensions attracted recently considerable attention\footnote{By now there exists a very extensive literature on the subject.
For reviews and further references see, e.g., Refs. \cite{Com,KL}.}.
Following the original work of Bardeen and Horowitz on the $d=4$ near horizon Kerr black hole \cite{bh}, such geometries are usually constructed by implementing a specific limit to
a given extreme black hole configuration (for a detailed discussion see, e.g., Ref. \cite{lmp}). In general, the limit yields a metric which can be interpreted as describing a complete vacuum spacetime on its own.
Its remarkable property is that the isometry group is extended to involve the conformal subgroup $SO(2,1)$, which motivates the holographic applications. The near horizon conformal symmetry of rotating black holes
also proved useful in the study of superconformal mechanics \cite{AG} and superintegrable models \cite{GNS}.

Denoting the temporal, radial, and azimuthal coordinates by $t$, $r$, and $\phi_i$, $i=1,\dots,n$, one can write the $SO(2,1)$--transformations in the form
\begin{align}\label{nhs}
&
t'=t+\alpha; &&  t'=t+\beta t, && r'=r-\beta r;
\nonumber\\[2pt]
&
t'=t+(t^2+\frac{1}{r^2}) \gamma, &&r'=r-2 tr\gamma, &&
\phi'_i=\phi_i-\frac{2}{r} \gamma,
\nonumber
\end{align}
where the infinitesimal parameters $\alpha$, $\beta$, and $\gamma$ correspond to the time translations, dilatations and special conformal transformations, respectively. Note that latitudinal coordinates remain inert under the action of the conformal group. Focusing on axially symmetric metrics which do not explicitly depend on the azimuthal angular variables $\phi_i$, the conformal invariants which typically enter the near horizon metrics read
\be\label{ci}
r^2 dt^2-\frac{dr^2}{r^2}, \qquad r dt+d \phi_i, \qquad d \phi_i-d\phi_j.
\nonumber
\ee
These are accompanied by coefficients which depend on latitudinal angular coordinates only.

One may wonder what happens if the logic is turned around and the conformal invariants are used to construct Ricci--flat metrics in diverse dimensions. It is known that not any
$SO(2,1)$--invariant geometry can be linked to a black hole predecessor (see, e.g., Ref. \cite{KL}). Yet, such geometries are definitely amenable to holographic applications and in some instances
they may provide useful insights into the structure of a black hole progenitor.

The goal of this work is to address this issue for the cases of four and five dimensions for which the vacuum Einstein equations reduce to a coupled set of ordinary differential equations. In four dimensions the analysis
can be carried out in full generality. In the next section we demonstrate that the $d=4$, $SO(2,1)$--invariant configuration corresponds to the near horizon Kerr--NUT black hole. In five dimensions the number of conformally invariant terms to be included into a metric grows notably. So in Sect. 3 we choose a specific ansatz whose structure is similar to the $d=5$ near horizon Myers--Perry black hole. A Ricci--flat metric which includes five arbitrary parameters is constructed. Setting two of them to vanish, one obtains a solution which seems to be a natural candidate to describe the $d=5$ near horizon Myers--Perry black hole with a NUT charge. In Sect. 4 we discuss
the results obtained in this work as well as possible further developments. Some technical issues related to the material presented in Sect. 3 are gathered in Appendix.

\vspace{0.5cm}

\noindent
{\bf 2. $d=4$ near horizon Kerr--NUT geometry}\\

\noindent
Given the conformal invariants, the most general $d=4$ metric invariant under the action of $SO(2,1)$ reads\footnote{Note that one more possible term of the type $p(\theta) d \theta (r dt+d \phi )$ can always be removed by redefining the variables $\theta$ and $\phi$.}
\be\label{metr}
ds^2=a(\theta)\left( r^2 dt^2-\frac{dr^2}{r^2}-d\theta^2\right)-b(\theta) {\left(r dt+d \phi \right)}^2,
\ee
where $\theta$ is the latitudinal angular variable.
The vacuum Einstein equations yield a coupled set of ordinary differential equations to determine the coefficients $a(\theta)$ and $b(\theta)$. A thorough investigation shows that they can be reduced to the
nonlinear ordinary differential equation for $a(\theta)$
\be\label{me}
4 a^2+2 a''(a -a'')+3 a'(a'+a^{(3)})=0,
\ee
while $b(\theta)$ is fixed provided $a(\theta)$ is known
\be\label{be}
b=\frac{4}{3} \left(a+a''\right)-\frac{a'^2}{a}.
\ee

It is worth mentioning that the near horizon extreme Kerr geometry which is characterized by the coefficients \cite{bh}
\be\label{bar}
a(\theta)=L_1 (1+\cos^2{\theta}), \qquad b(\theta)=\frac{4 L_1 \sin^2{\theta}}{1+\cos^2{\theta}},
\ee
where $L_1$ is a constant related to the rotation parameter, does provide a particular solution to Eqs. (\ref{me}) and (\ref{be}). As (\ref{me}) is the third order ordinary differential equation, its general solution involves three constants of integration. It is easy to verify that shifting $\theta$ in (\ref{bar}) by a constant one obtains a new solution to (\ref{me}) and (\ref{be}). However, the new parameter is physically irrelevant as it does not alter the metric. It is then natural to expect that the Bardeen--Horowitz solution (\ref{bar}) can be extended to include one more arbitrary parameter, the latter to be identified with a NUT charge.

In order to solve (\ref{me}) in full generality, we first note that it is homogeneous in $a(\theta)$ and its derivatives. This justifies the substitution
\be\label{sub}
a(\theta)=e^{q(\theta)},
\ee
which is consistent with the signature of the metric chosen and gives a simpler third order differential equation for $q(\theta)$. As the latter does not involve $\theta$ and $q(\theta)$ explicitly, the two consecutive substitutions
\be
q'(\theta)=p(\theta), \qquad p'(\theta)=s(p(\theta))
\ee
reduce it to the first order equation for $s(p)$. The latter can be further simplified by introducing the new variable
\be
y=p^2,
\ee
which yields
\be\label{eqs}
(4+5 y+y^2)+s \left(2+5 y-2 s+6 y s'\right)=0,
\ee
where $s'(y)=\frac{d s(y)}{d y}$. This is a variant of the Abel equation which can be explicitly solved in some exceptional cases only. Representing $s(y)$ in the form
\be
s(y)=-(1+y)+u(y)
\ee
where $-(1+y)$ is a particular solution to (\ref{eqs}), one converts (\ref{eqs}) into the equation for $u(y)$
\be\label{equ}
-2 u^2-6 y(1+y) u'+3 u(2+y+2 y u')=0
\ee
with the coefficients in front of $u(y)$ and $u'(y)$ being quadratic polynomials in $y$. The latter fact prompts one to search for
the general solution to Eq. (\ref{equ}) in the parametric form
\be
y=w(z), \qquad u(y)=z w(z),
\ee
which ultimately yields
\be
w(z)=\frac{-9 + 6 z + 2 c_1 \sqrt{-3 + 4 z}}{2 z^2},
\ee
where $c_1$ is a constant of integration.

When returning back to $a(\theta)$, it proves technically convenient to keep the variable $z$ explicit until the very last step
\be
a(z)=\frac{c_3 z}{-3 + 4 z}, \qquad \theta(z)=c_2-\arctan{\frac{-9 + 2 c_1 \sqrt{-3 + 4 z}}{
   3 \sqrt{-9 + 4 c_1 \sqrt{-3 + 4 z} + 3 (-3 + 4 z)}}},
\ee
where $c_2$ and $c_3$ are constants of integration. Solving the rightmost equation for $z$ and removing $c_2$ by redefining $\theta$,
one finally gets
\be\label{eqa}
a(\theta)=L_1 (1+\cos^2{\theta})+L_2 \cos{\theta},
\ee
where $L_1$ and $L_2$ are arbitrary parameters. The form of the function $b(\theta)$ follows from (\ref{be})
\be\label{eqb}
b(\theta)=\frac{(4 L_1^2 - L_2^2) \sin^2{\theta}}{a(\theta)}.
\ee
Note that the resulting metric (\ref{metr}) has the Lorentzian signature provided
\be
4 L_1^2>L_2^2,
\ee
while for $L_1=0$ the solution is of the ultrahyperbolic signature $(2,2)$.

Remarkably enough, the metric which we constructed by providing the general solution to Eq. (\ref{me}) precisely coincides with the near horizon limit of the extreme Kerr--NUT black hole \cite{Gh}.
$L_1$ can be linked to the rotation parameter, while $L_2$ represents a NUT charge. We thus conclude that in four dimensions the $SO(2,1)$--invariance allows one to unambiguously fix the NUT--extension of the near horizon Kerr geometry.

\vspace{0.5cm}

\noindent
{\bf 3. NUT--extension of $d=5$ near horizon Myers--Perry geometry }\\

\noindent
In five dimensions the number of $SO(2,1)$--invariant terms to be included into a metric grows notably. So we choose a specific ansatz
\bea\label{5dm}
&&
ds^2=a(\theta)\left(r^2 dt^2-\frac{dr^2}{r^2}-d\theta^2\right)-b(\theta) {(r dt+d \phi_1)}^2 -c(\theta){(r dt+d \phi_2)}^2+
\nonumber\\[2pt]
&&
\quad \qquad +d(\theta) {(d \phi_1-d \phi_2)}^2,
\eea
with $a(\theta)$, $b(\theta)$, $c(\theta)$, $d(\theta)$ to be determined,
whose structure is similar to the $d=5$ near horizon Myers--Perry black hole \cite{lmp}~\footnote{As compared to the notation in \cite{lmp}, we redefined the latitudinal angular variable $2\theta~\rightarrow~\theta-\frac{\pi}{2}$ and omitted the overall factor $\frac 12(a_1+a_2)$.}
\bea\label{MP}
&&
ds^2=\alpha(\theta) \left(r^2 dt^2-\frac{dr^2}{r^2}-d\theta^2\right)-\frac{a_2(1-\sin{\theta}) (2a_1+\alpha(\theta) )}{\alpha(\theta) } {(r dt+d \phi_1)}^2 -
\nonumber\\[2pt]
&&
\quad \qquad -\frac{a_1(1+\sin{\theta}) (2 a_2+\alpha(\theta) )}{\alpha(\theta) } {(r dt+d \phi_2)}^2+\frac{a_1 a_2 \cos^2{\theta}}{\alpha(\theta)} {(d \phi_1-d \phi_2)}^2,
\nonumber\\[2pt]
&&
\alpha(\theta)=a_1+a_2+(a_1-a_2) \sin{\theta},
\eea
where $a_1$ and $a_2$ represent the rotation parameters.Throughout this work we consider the case of nonzero and unequal rotation parameters.

A careful analysis of components of the Ricci tensor constructed from the metric (\ref{5dm}) shows that $R_{t \phi_1}$, $R_{t \phi_2}$, $R_{rr}$, $R_{\phi_1 \phi_2}$ produce the coupled set of second order ordinary differential equations
\bea\label{eq1}
&&
a''=\frac{(2 a(b+c)-4 a^2+a'^2)g- a a' g'}{2 a g},
\\[2pt]\label{eq2}
&&
b''=\frac{-(2b(b+c)+a' b')g+a b' g'+2 ab(c'd'-b'c'+b'd')}{2ag},
\\[2pt]\label{eq3}
&&
c''=\frac{-(2c(b+c)+a' c')g+a c' g'+2 ac(c'd'-b'c'+b'd')}{2ag},
\\[2pt]\label{eq4}
&&
d''=\frac{-(2bc+a' d')g+a d' g'+2 ad(c'd'-b'c'+b'd')}{2ag},
\eea
where we denoted\footnote{It is worth mentioning that $g$ is proportional to the determinant of the metric
$\mbox{det} (g_{ij})=4 a^3 g$.}
\be\label{eqQ}
g=bc - d(b+c),
\ee
while $R_{\theta \theta}$ yields the compatibility condition
\bea\label{eq6}
&&
c'd'-b'c'+b'd'=\frac{(4 a^2-a(b+c)+a'^2)g+2 a a' g'}{a^2}.
\eea
One more compatibility condition comes from the definition (\ref{eqQ}) and Eqs. (\ref{eq1})--(\ref{eq4})
\bea\label{eq5}
&&
g''=\frac{-(2(b+c)g+a'g')g+a g'^2}{2 a g}.
\eea
Other components of the Ricci tensor prove to vanish identically, provided Eqs. (\ref{eq1})--(\ref{eq4}) and Eq. (\ref{eq6}) hold.

The system of ordinary differential equations exposed above can be solved in full generality. Gathering technical details in Appendix, we display below a solution which, in our opinion, seems to be a natural candidate
to describe the $d=5$ near horizon Myers--Perry black hole with a NUT charge
\bea\label{finres}
&&
a(\theta)=L_1+L_2 \sin{\theta}+L_3 \sin^2{\theta}, \qquad \quad d(\theta)=\frac{a(\theta)b(\theta)c(\theta)-N \cos^2{\theta}}{a(\theta)(b(\theta)+c(\theta))},
\nonumber\\[4pt]
&&
b(\theta)=\frac{(L_1 - L_2) (2 L_1 + L_2 -
   2(L_1  - L_3 (1+2 L_1/L_2))\sin{\theta}-L_2 \sin^2{\theta}) }{2 a(\theta)},
\nonumber\\[4pt]
&&
c(\theta)=\frac{2 L_1 (L_1 + L_3)-L_2^2 +
 2 L_2 (L_1 - L_3) \sin{\theta} + (L_2^2 - 2 L_3 (L_1 + L_3)) \sin^2{\theta}}{a(\theta)}-b(\theta),
\nonumber\\[4pt]
&&
N =\frac{ {(L_1 - L_2)}^2 {(L_1 L_2(L_1 +  L_2) - 2 L_3 L_1^2  -L_3^2 (2 L_1 + L_2))}^2}{
2 L_2^2 (L_1 - L_3) (L_1^2 - L_2^2 + L_3(2 L_1  + L_3))},
\eea
where $L_1$, $L_2$ and $L_3$ are constants. A more detailed form of the function $d(\theta)$ is given in Appendix.

As follows from Eq. (\ref{MP}), $L_1$ and $L_2$ can be linked to the rotation parameters via
\be
L_1=a_1+a_2, \qquad L_2=a_1-a_2.
\ee
By analogy with the $d=4$ case, it seems natural to interpret $L_3$ as a NUT charge. In particular, in the limit $L_3 \to 0$ the solution (\ref{finres}) reduces to (\ref{MP}). Note that the metric has the Lorentzian signature provided
\be
(L_1 - L_3) (L_1^2 - L_2^2 + 2 L_1 L_3 + L_3^2)>0.
\ee

As is shown in the Appendix, the functions $b$,$c$ and $d$ in (\ref{finres}) can be deformed to include two more arbitrary parameters in such a way that the resulting metric (\ref{5dm}) still provides a solution to the vacuum Einstein equations. A geometrical or physical interpretation of the extra parameters remains a challenge.

\vskip 0.5cm
\noindent
{\bf 4. Discussion}\\

\noindent
To summarize, in this work we employed the invariants of the conformal group $SO(2,1)$ so as to construct Ricci--flat metrics in four and five dimensions. Our consideration was primarily concerned with $d=4$ and $d=5$ because in these cases
the vacuum Einstein equations reduced a coupled set of ordinary differential equations which could be analyzed in full generality. In four dimensions the resulting metric reproduced the near horizon Kerr--NUT
black hole \cite{Gh}. To the best of our knowledge, the five--dimensional metric presented in Sect. 3 is new. It involves five arbitrary parameters. Setting two of them to vanish, one obtains a natural candidate to describe the $d=5$ near
horizon Myers--Perry black hole with a NUT charge.

The NUT--charged rotating black hole geometries in arbitrary dimension have been constructed in Refs. \cite{CLP,CLP1}. They were built by appropriately equating rotation parameters and making use of a special coordinate system. In particular, according to the results in \cite{CLP,CLP1}, in five dimensions a NUT charge is bogus as it can be removed by redefining the variables. This implies that our metric in Sect. 3 cannot be obtained as the near horizon limit of that in \cite{CLP,CLP1}.

In odd dimensions the NUT charges enter the metric in \cite{CLP1} as additive constants. Note, however, that in the absence of rotation parameters 
NUT charges typically accompany terms involving the latitudinal angular variables and they are not just additive constants (see, e.g., the construction in Ref. \cite{MS}). 
In this regard the $d=4$ and $d=5$ metrics constructed above are universal and involve a NUT charge in a uniform way.

Turning to possible further developments, it is worth mentioning that in Ref. \cite{LMP} yet even more general five--dimensional solution has been constructed which involves one extra parameter
over and above the rotation parameters characterizing the Myers--Perry black hole. It would be interesting to study the near horizon limit of the metric in \cite{LMP} and confront it with that in Sect. 3. In this regard the important thing to understand is how the coordinate systems used in \cite{LMP} is related to that in this work.
A generalization to $d>5$, including the case of a nonvanishing cosmological constant, is an important open problem. In this case, in order to solve the vacuum Einstein equations, one has to deal with a coupled set of partial differential equations which are technically much more difficult to solve. Finally, it would be interesting to construct integrable systems associated with NUT--charged near horizon black hole geometries in the spirit of \cite{GNS}.

\vskip 0.5cm
\noindent
{\bf Acknowledgements}\\

\noindent
This work was supported by the MSE program Nauka under the project 3.825.2014/K and the RFBR grant 15-52-05022.

\vskip 0.5cm
\noindent
{\bf Appendix}\\

\noindent
In this Appendix we discuss some technical issues involved in the construction of the solution (\ref{finres}) to the
system of ordinary differential equations (\ref{eq1})--(\ref{eq4}) and the compatibility conditions (\ref{eq6}), (\ref{eq5}).

Multiplying Eqs. (\ref{eq1}) and (\ref{eq5}) by $g$ and $a$, respectively, and taking the sum, one gets the simple differential equation
\be
(ag)''=\frac{{(ag)'}^2}{2ag}-2ag
\nonumber
\ee
whose general solution reads
\be\label{eqQQ}
a g=c_1 \cos^2{(\theta+c_2)},
\nonumber
\ee
where $c_1$ and $c_2$ are constants of integration. In what follows we disregard $c_2$ as it can be eliminated by redefining $\theta$. Taking into account the definition (\ref{eqQ}), one concludes that one of the functions
$a$, $b$, $c$, or $d$ can be algebraically expressed in terms of the others. For definiteness, we choose
\be
d=\frac{abc-c_1 \cos^2{\theta}}{a(b+c)}.
\nonumber
\ee

One more algebraic relation is obtained by substituting the solution for the product $ag$ into Eq. (\ref{eq1}), which yields
\be\label{bps}
b+c=a''+2a- a' \left(\tan{\theta}+\frac{a'}{a}\right).
\nonumber
\ee
Taking into account this expression and computing the sum of Eqs. (\ref{eq2}) and (\ref{eq3}), one gets the fourth order ordinary differential equation to fix $a(\theta)$
\bea\label{fo}
&&
a\left(3a'\cos{2\theta} \sec^2{\theta} \tan{\theta}+a''(7-3\sec^2{\theta}) +a^{(4)}\right)-
\nonumber\\[2pt]
&&
-a'\left(a'(-2+3\sec^2{\theta})+3a''\tan{\theta} +a^{(3)} \right)=0.
\nonumber
\eea
In order to simplify it,
we introduce the new variable
\be
y=\sin{\theta},
\nonumber
\ee
represent $a$ in the form
\be
a(\theta)=e^{q(y(\theta))},
\nonumber
\ee
and implement two consecutive substitutions
\be\label{Eqp}
q'=p, \qquad p^3+3 p p'+p''=u,
\nonumber
\ee
where the prime denotes the derivative with respect to $y$. Then the fourth order equation for $a$ reduces to the first order equation for $u$
\be
(1-y^2) u'-6 y u=0,
\nonumber
\ee
which has the simple solution
\be
u(y)=\frac{u_0}{{(y^2-1)}^3},
\nonumber
\ee
where $u_0$ is a constant of integration.
The general solution to the equation $p^3+3 p p'+p''=u$ is obtained by making use of the substitution
\be\label{ppp}
p=\frac{w'}{w},
\nonumber
\ee
which raises the order of the equation by one
\be\label{eqw}
\frac{w^{(3)}}{w}=\frac{u_0}{{(y^2-1)}^3}.
\nonumber
\ee
Taking into account the relations $q'=p$ and $p=\frac{w'}{w}$, one concludes that $a(\theta)$ coincides with $w(y(\theta))$ up to a constant factor. The integration of the equation for $w$ then gives
\be
a(y)= (y^2-1) \left( a_1 {\left(\frac{y-1}{y+1}\right)}^{a_4}+a_2 {\left(\frac{y+1}{y-1}\right)}^{\frac{a_4+\sqrt{4-3a_4^2}}{2}}+a_3  {\left(\frac{y+1}{y-1}\right)}^{\frac{a_4-\sqrt{4-3a_4^2}}{2}}\right),
\nonumber
\ee
where $y=\sin{\theta}$ and $a_1$, $a_2$, $a_3$, $a_4$ are constants of integration. 

Since in this work we are primarily concerned with the construction of a NUT--extension of the $d=5$ near horizon Myers--Perry geometry and
$a(y)$ in the preceding formula is a transcendental function, in what follows we choose $a_4$ to take one of the three integer values $-1$, $0$, or $1$, which yield\footnote{It is likely that irrational values of $a_4$ lead to trivial solutions. For example, choosing $a_1=a_3=0$, $a_4=\frac 14 (1-\sqrt{13})$, one finds $a(\theta)=a_2 \cos{\theta}(1+\sin{\theta})$. It is straightforward to verify that this form of $a(\theta)$ implies $b(\theta)=c(\theta)=0$ which lead to the divergent metric. }
\be\label{afin}
a(\theta)=L_1+L_2 \sin{\theta}+L_3 \sin^2{\theta},
\nonumber
\ee
where $L_1$, $L_2$ and $L_3$ are arbitrary parameters. A comparison with (\ref{MP}) shows that the first two constants can be linked to the rotation parameters via $L_1=a_1+a_2$ and $L_2=a_1-a_2$ , while $L_3$ can be interpreted as a NUT charge.

Given the explicit form of the function $a$, one can immediately compute $g$, $(b+c)$, and $c'd'-b'c'+b'd'$. At this stage, the linear ordinary differential equation (\ref{eq2}) can be integrated to yield
\be\label{finb}
b(\theta)=\frac{b_1 (L_2 + 2 L_3 \sin{\theta}) + b_2 (2 L_1 \sin{\theta} + L_2 \sin^2{\theta}) }{a(\theta)},
\nonumber
\ee
where $b_1$, $b_2$ are constants of integration and the functions they accompany represent the two linearly independent solutions to Eq. (\ref{eq2}). Because in this work we are concerned with the construction of a NUT--deformation
of the $d=5$ near horizon Myers--Perry geometry, we choose to fix $b_1$ and $b_2$ in such a way that the resulting metric reduces to (\ref{MP}) in the limit $L_3\to 0$. This gives
\be\label{b1b2}
b_1=\frac{1}{2L_2}\left(2 L_1^2-L_2(L_1+L_2) \right), \qquad b_2=\frac 12 (L_2-L_1),
\nonumber
\ee
which ultimately yield $b(\theta)$ of the form
\be
b(\theta)=\frac{(L_1 - L_2) (2 L_1 + L_2 -
   2(L_1  - L_3 (1+2 L_1/L_2))\sin{\theta}-L_2 \sin^2{\theta}) }{2 a(\theta)}.
\nonumber
\ee

Because the equation (\ref{eq3}) for $c$ has exactly the same form as that for $b$, its solution has a similar structure
\be\label{finb}
c(\theta)=\frac{C_1 (L_2 + 2 L_3 \sin{\theta}) + C_2 (2 L_1 \sin{\theta} + L_2 \sin^2{\theta}) }{a(\theta)},
\nonumber
\ee
where $C_1$ and $C_2$ are constants of integration. 
Taking into account the relation $b+c=a''+2a- a' \left(\tan{\theta}+\frac{a'}{a}\right)$, one can relate $C_1$ and $C_2$ to $b_1$, $b_2$, $L_1$, $L_2$ and $L_3$. The ultimate result reads
\bea\label{finc}
&&
c(\theta)=\frac{L_1 L_2 - L_2^2+2 L_1^2+ 4 L_3 L_1  +
 2 (L_1^2+ L_1 L_2 +L_3(L_1-L_2-2 L_1^2/L_2)) \sin{\theta}}{2 (L_1+L_2 \sin{\theta}+L_3 \sin^2{\theta})}+
 \nonumber\\[2pt]
 &&
 \qquad \quad
+ \frac{
 (L_2 (L_1 + L_2) - 4  L_3(L_1 + L_3)) \sin^2{\theta}}{2 (L_1+L_2 \sin{\theta}+L_3 \sin^2{\theta})}.
 \nonumber
\eea

It remains to fix the function $d$. The simplest way to solve Eq. (\ref{eq4}) is to start with the ansatz
\be
d(\theta)=\frac{d_1+d_2 \sin{\theta}+d_3 \sin^2{\theta}}{a(\theta)}
\ee
and then determine the constants $d_1$, $d_2$ and $d_3$ from Eq. (\ref{eq4}) and the compatibility conditions (\ref{eqQ}), (\ref{eq6}). This fixes $d$ unambiguously. In order to facilitate the comparison with (\ref{MP}),  below we represent the constants as power series in $L_3$
\bea
&&
d_1=(L_1-L_2)\left(L_1 (L_1 - L_2) L_2^2 {(L_1 + L_2)}^2+L_3 L_2 (L_1 + L_2) (4 L_1^3 + L_1^2 L_2 + L_2^3)-
\right.
\nonumber\\[2pt]
&& \quad \quad
\left.
-L_3^2 L_1 (2 L_1 + L_2) (2 L_1^2 - 3 L_1 L_2 + 3 L_2^2)-L_3^3 (2 L_1 + L_2) (2 L_1^2 - L_1 L_2 + L_2^2)\right)/d_4,
\nonumber\\[4pt]
&&
d_2=L_3(L_1 - L_2)\left(-4  L_1 L_2^3 (L_1 + L_2)+2 L_3 L_2 (L_1 + L_2) (4 L_1^2 + 2 L_1 L_2 + L_2^2)-
\right.
\nonumber\\[2pt]
&& \quad \quad
\left.
-8 L_3^2 L_1^2 L_2-4 L_3^3 L_2 (2 L_1+  L_2)\right)/d_4,
\nonumber\\[4pt]
&&
d_3=-(L_1-L_2)\left(L_1 (L_1 - L_2) L_2^2 {(L_1 + L_2)}^2 -L_3 L_2 (L_1 + L_2) (4 L_1^3 - 5 L_1^2 L_2 - L_2^3)+
\right.
\nonumber\\[2pt]
&& \quad \quad
\left.
+L_3^2 L_1 (4 L_1^3 - 4 L_1^2 L_2 - L_1 L_2^2 - L_2^3)+L_3^3 (4 L_1^3 - 3 L_1 L_2^2 - 3 L_2^3)\right)/d_4,
\nonumber
\eea
where $d_4$ reads
\be
d_4=4 L_2^2 \left( L_1 (L_1^2 - L_2^2) +  L_3 (L_1^2 + L_2^2) -  L_3^2 L_1   -
  L_3^3\right).
  \nonumber
\ee
In the process one also determines the constant $c_1$ which enters the expression for $d$ exposed above
\be
c_1 =\frac{ {(L_1 - L_2)}^2 {(L_1 L_2(L_1 +  L_2) - 2 L_3 L_1^2  -L_3^2 (2 L_1 + L_2))}^2}{
2 L_2^2 (L_1 - L_3) (L_1^2 - L_2^2 + L_3(2 L_1  + L_3))}.
\nonumber
\ee

In our previous discussion we have fixed the arbitrary constants $b_1$ and $b_2$ entering the function $b(\theta)$ so as to conform to the $d=5$ near horizon Meyers--Perry metric which shows up in the limit $L_3\to 0$. Leaving them arbitrary, one can construct a two--parametric deformation of
the solution (\ref{finres})
\bea
&&
\tilde b(\theta)=b(\theta)+\frac{P_1 L_2 + 2 (P_2 L_1 + P_1 L_3) \sin{\theta} + P_2 L_2 \sin^2 {\theta}}{a(\theta)},
\nonumber\\[2pt]
&&
\tilde c(\theta)=c(\theta)-\frac{P_1 L_2 + 2 (P_2 L_1 + P_1 L_3) \sin{\theta} + P_2 L_2 \sin^2 {\theta}}{a(\theta)},
\nonumber\\[2pt]
&&
\tilde d(\theta)=d(\theta)+\frac{\tilde d_1+\tilde d_2 \sin{\theta}+\tilde d_3 \sin^2{\theta}}{a(\theta)},
\nonumber
\eea
where $P_1$ and $P_2$ are the new parameters. The function $a(\theta)$ maintains its form (\ref{finres}), while the constants $\tilde d_1$, $\tilde d_2$, $\tilde d_3$ are expressed via $L_1$, $L_2$, $L_3$, $P_1$ and $P_2$ as follows:
\bea
&&
\tilde d_1=(-2 P_2 L_1 (-L_1^3 L_2 + (L_1-P_1) L_2^3 +
     2 L_1 (L_1^2 + P_1 L_2 -
        L_1 L_2) L_3 +
\nonumber
\\[2pt]
&&
\quad +(2 P_1 L_2 + (L_1 - L_2) (2 L_1 + L_2)) L_3^2) -
  P_1 (P_1 L_1 L_2^3 - 2 L_2 (L_1^3 + 2 L_1^2 L_2 + P_1 L_2^2 -
\nonumber
\\[2pt]
&&
\quad -L_1 L_2^2) L_3 +2 L_1 (2 L_1^2 - 2 L_1 L_2 + L_2 (P_1 + L_2)) L_3^2 +
     2 (2 L_1^2 + P_1 L_2 - L_1 L_2) L_3^3) -
\nonumber
\\[2pt]
&&
\quad
-P_2^2 L_1 L_2 (-L_2^2 + 2 L_1 (L_1 + L_3)))/\tilde d_4,
\nonumber
\\[2pt]
&&
\tilde d_2=L_2(P_1^2 L_2 (L_2^2 - 4 L_1 L_3) +
  2 P_1 (L_1^2 (L_1-2 P_2) L_2 + (P_2 - L_1) L_2^3 +
     2 L_1^2 (L_2-
\nonumber
\\[2pt]
&&
\quad
-L_1) L_3 +(2 L_1^2 + L_1 L_2 + L_2 (L_2-2 P_2)) L_3^2 -
     2 L_1 L_3^3 - 2 L_3^4) +
  P_2 (4 L_1^2 (L_2 -
\nonumber
\\[2pt]
&&
\quad
- 2 L_3) L_3
+2 L_1^3 (L_2 + 2 L_3) +
     L_2^2 (P_2 L_2 + 2 L_3^2) -
     2 L_1 (L_2^3 + L_2 (2 P_2 - L_3) L_3 + 2 L_3^3)))/\tilde d_4,
\nonumber
\\[2pt]
&&
\tilde d_3=(-P_2^2 L_2 (2 L_1^3 - 2 L_1 L_2^2 + 2 L_1^2 L_3 + L_2^2 L_3) -
  P_1 L_3 (-P_1 L_2^3 - 2 L_1^3 (L_2 - 2 L_3) +
\nonumber
\\[2pt]
&&
\quad
+2 (P_1 - L_2) L_2 L_3^2 +
     4 L_1^2 L_3 (L_3-L_2) + 2 L_1 L_2 (L_2^2 + (P_1 - L_3) L_3)) -
\nonumber
\\[2pt]
&&
\quad
-2 P_2 (-L_1^4 (L_2 - 2 L_3) + 2 P_1 L_1 L_2 L_3^2 + 2 L_1^3 L_3 (L_3-L_2) +
     L_1^2 L_2 (L_2^2 - 2 L_2 L_3 +
\nonumber
\\[2pt]
&&
\quad
+(2 P_1 - L_3) L_3)
+L_2^2 L_3 (-P_1 L_2 + L_3^2)))/\tilde d_4,
\nonumber
\\[2pt]
&&
\tilde d_4=2 L_2 (L_1 - L_3) (L_1 - L_2 + L_3) (L_1 + L_2 + L_3).
\nonumber
\eea
It is straightforward to verify that the modified functions do provide a solution to the vacuum Einstein equations
which reduces to (\ref{finres}) in the limit $P_1\to 0$, $P_2\to0$. A geometrical or physical interpretation of the extra parameters remains a challenge.

\end{document}